\documentclass[acmlarge,screen]{acmart}
\usepackage{tabularx} 
\usepackage{makecell}
\usepackage{arydshln}
\usepackage{hyperref}
\AtBeginDocument{%
  \providecommand\BibTeX{{%
    \normalfont B\kern-0.5em{\scshape i\kern-0.25em b}\kern-0.8em\TeX}}}

\setcopyright{acmcopyright}
\copyrightyear{2024}
\acmYear{2024}
\acmDOI{XXXXXXX.XXXXXXX}

\acmJournal{IMWUT}
\acmVolume{0}
\acmNumber{0}
\acmArticle{111}
\acmMonth{8}

\begin{document}

% \title{Aging Dementia Healthcare Research in HCI - A Scoping Review }
\title{Bridging the Gap: Advancements in Technology to Support Dementia Care – A Scoping Review}

\author{Yong Ma}
\affiliation{%
  \institution{University of Bergen}
  \city{Bergen}
  \country{Norway}}
  \email{yong.ma@uib.no}

\author{Oda Elise Nordberg}
\affiliation{%
\institution{University of Bergen}
  \city{Bergen}
  \country{Norway}}
\email{Oda.Nordberg@uib.no}

\author{Jessica Hubbers}
\affiliation{%
\institution{Helse Fonna}
  \city{Haugesund}
  \country{Norway}}
\email{jessica.hubbers@helse-fonna.no}

\author{Yuchong Zhang}
% \orcid{0000-0003-1804-6296}
\email{yuchongz@kth.se}
\affiliation{%
  \institution{KTH Royal Institute of Technology}
  \city{Stockholm}
  \country{Sweden}
}

\author{Arvid Rongve}
\affiliation{%
\institution{Helse Fonna}
  \city{Haugesund}
  \country{Norway}}
\email{arvid.rongve@helse-fonna.no }
\additionalaffiliation{%
\institution{University of Bergen}
  \city{Bergen}
  \country{Norway}}
\email{arvid.rongve@uib.no}

\author{Miroslav Bachinski}
\affiliation{%
\institution{University of Bergen}
  \city{Bergen}
  \country{Norway}}
\email{miroslav.bachinski@uib.no}

\author{Morten Fjeld}
% \orcid{0000-0002-9562-5147}
\affiliation{%
\institution{University of Bergen}
  \city{Bergen}
  \country{Norway}}
\email{morten.fjeld@uib.no}
\additionalaffiliation{%
\institution{Chalmers University of Technology}
  \city{Gothenburg}
  \country{Sweden}}
\email{fjeld@chalmers.se}

\renewcommand{\shortauthors}{Ma et al.}

\begin{abstract}
Dementia has serious consequences for the daily life of the person affected due to the decline in the their cognitive, behavioral and functional abilities. Caring for people living with dementia can be challenging and distressing. Innovative solutions are becoming essential to enrich the lives of those impacted and alleviate caregiver burdens. This scoping review, spanning literature from 2010 to July 2023 in the field of Human-Computer Interaction (HCI), offers a comprehensive look at how interactive technology contributes to dementia care. Emphasizing technology's role in addressing the unique needs of people with dementia (PwD) and their caregivers, this review encompasses assistive devices, mobile applications, sensors, and GPS tracking. Delving into challenges encountered in clinical and home-care settings, it succinctly outlines the influence of cutting-edge technologies, such as wearables, virtual reality, robots, and artificial intelligence, in supporting individuals with dementia and their caregivers. We categorize current dementia-related technologies into six groups based on their intended use and function: 1) daily life monitoring, 2) daily life support, 3) social interaction and communication, 4) well-being enhancement, 5) cognitive support, and 6) caregiver support. 
% Furthermore, we delves into the future trajectory of dementia care, envisioning a landscape where robot technologies, driven by artificial intelligence and human-robot interaction, may play a significant role, potentially reshaping the landscape of dementia care.
\end{abstract}






\maketitle

\section{Introduction}
Dementia is a progressive neurological condition that is characterized by cognitive decline, behavioral changes, and functional impairments~\cite{dening2015dementia}. The gradual loss of memory and cognitive, emotional, and social abilities adversely affects daily activities~\cite{potkin2002abc}, and diminishes the overall quality of life ~\cite{andersen2004ability}. Although dementia is closely linked to advancing age, it is important to emphasize that age acts as a risk factor rather than a causal factor for dementia ~\cite{dening2015dementia}. Dementia has serious consequences for the daily lives of the person affected, as well as their informal caregivers. As the global population ages, the two main challenges associated with dementia are early-stage detection and dementia care. 
% With the advancement of Artificial Intelligence (AI) and wearable technology, the early-stage diagnosis of dementia can be achieved through the analysis of patients' brain imaging, speech patterns, facial expression, movement patterns, sleeping behavior, etc~\cite{li2022applications}.
% With the progress of Artificial Intelligence (AI) and wearable technology, early-stage dementia diagnosis can be accomplished through the analysis of patients' brain imaging, speech features, facial expression, movement patterns, sleeping behavior, etc~\cite{li2022applications}. 
% Caring for Patients with Dementia (PwD) has become a main increasingly challenging issue as the prevalence of dementia has increased. 
With the advancements in artificial intelligence (AI) and the proliferation of wearable technology, the early-stage diagnosis of dementia can now be achieved through the analysis of various patient data, including brain imaging, speech characteristics, facial expressions, movement patterns, sleeping behavior, and more~\cite{li2022applications}. 
% As the prevalence of dementia continues to rise, caring for individuals with dementia (PwD) has become an increasingly complex and pressing challenge, putting a growing burden on the healthcare system and caregivers.
% % Most often, those who provide care for PwD are simultaneously their spouses, partners, or adult children. In many cases, these caregivers have limited prior experience in healthcare and would greatly benefit from access to training and support resources.
% Although formal or professional caregivers can provide the good healthcare, long-term caring and highly cost may become the burden on the normal family.
% Generally, those who provide care for PwD are simultaneously their spouses, partners, or adult children. In many cases, these caregivers have limited prior experience in healthcare and would greatly benefit from access to training and support resources.
As the prevalence of dementia continues its upward trajectory, caring for People with dementia (PwD) has become an increasingly complex and pressing challenge. This puts a mounting burden not only on the healthcare system but also on the shoulders of devoted caregivers. 
% Generally, professional dementia caregivers can provide holistic care to PwD in need, which including disease progression monitoring, medication management, falls prevention, activation and engagement, emotional support and stress reduction~\cite{stephan2018barriers,hubbers2021dementia}.
In general, professional caregivers are able to provide personal care to PwD in need, which can include disease progression monitoring, medication management, fall prevention, activation and engagement, emotional support, and stress reduction~\cite{stephan2018barriers,hubbers2021dementia}. Olsen et al.~\cite{Olsen2016} found that, beyond economic and healthcare political reasons, PwD in Norway experienced significantly improved quality of life when living at home compared to those in nursing homes, regardless of the severity of their dementia. While formal or professional caregivers can deliver high-quality healthcare, the long-term commitment and associated costs can become an unwelcome strain on ordinary families. In many cases, those assuming the role of caregivers for PwD are simultaneously their spouses, partners, or adult children. It is noteworthy that these caregivers often possess limited prior experience in the realm of healthcare, rendering them particularly receptive to the benefits derived from accessible training and support resources.

% Caring for individuals living with dementia can indeed be a challenging and emotionally taxing responsibility, particularly when these caregivers are simultaneously the spouses, partners, or adult children of those affected. Often, these caregivers have little prior experience in providing care, highlighting the critical need for training and support. This review paper aims to offer a comprehensive and systematic overview of how Human-Computer Interaction (HCI) has positively influenced dementia care. One promising approach is the utilization of Virtual Reality (VR) as a training strategy, which has shown potential in enhancing the skills of dementia caregivers.
Dementia care has been shown to benefit from assistive technologies, such as monitoring and security technologies, daily living support services, cognitive-focused therapies, and so on~\cite{pappada2021assistive}. 
% For some instances, wearable devices with multiple sensors can provide PwD with daily life monitoring~\cite{yang2021multimodal}, and mobile-health applications can give PwD with healthcare support~\cite{yousaf2019mobile}. Furthermore, daily activity data visualization for PwD~\cite{lippe2021variables}, can provide valuable insights into behavior and care patterns, and these information can benefit their caregivers. 
% For instance, wearable devices equipped with an array of sensors, as elucidated by Yang et al.~\cite{yang2021multimodal}, have emerged as a pivotal tool in providing continuous daily life monitoring for people with dementia (PwD). 
For instance, wearable devices equipped with an array of sensors, as elucidated by Yang et al.~\cite{yang2021multimodal}, can provide continuous daily life monitoring for people with dementia (PwD). 
Simultaneously, mobile health applications, as discussed by Yousaf et al.~\cite{yousaf2019mobile}, extend invaluable healthcare support, enhancing the overall well-being of PwD.
Moreover, the visualization of daily activity data for PwD, as explored by Lippe et al.~\cite{lippe2021variables}, offers profound insights into their behavior and care patterns. This information not only enriches the understanding of their needs but also proves invaluable to their caregivers, enabling them to provide more tailored and effective support.
% Additionally, interactive technology, such as entertainment system that can develop an engaging multimedia activity for PwD to use independently~\cite{alm2007interactive}, can enhance greatly enhance the overall experience of PwD.
Innovative interactive technologies, such as entertainment systems designed to facilitate engaging multimedia activities for PwD, as exemplified by Alm et al.~\cite{alm2007interactive}, have the power to significantly enhance the overall experience of individuals living with dementia. These technologies foster independence and joy, contributing to a higher quality of life for those affected.

Even though it has been proven that technology can assist PwD in several aspects, there are some debates in the field of HCI concerning ethical issues of technologies for PwD. As with many other conditions, dementia is often viewed through the medical lens as a deficit and a problem that can be "solved" by the use of technology. This notion has sparked debate, for example by the critical dementia position, where researchers recognize the stigmatization and disempowerment people with dementia often face in our society ~\cite{lazar2017critical}. According to this view, when developing technology for PwD the focus is mainly of "prescribing solutions" to their deficits. However, researchers from the critical dementia position argue that researchers should rather have ongoing communication with the PwD and offer tailored technologies. When developing technologies for PwD, Fabricatore et al. ~\cite{Fabricatore2020} mentions several questions researchers should ask themselves, for example who the intended user of the solution is (PwD vs. their caretaker), how it should be used (e.g. as a monitoring or a retaining tool), why it should be developed (what challenges it supports), and how it has been adjusted to PwD. By doing so, it might be possible to consider technology for PwD in a more nuanced manner. Even though there are some critical debates in the field, they seem to mainly revolve around how technologies should be adapted to PwD and what challenges it can be useful for. 
 
% Based on these technological breakthroughs and traditional dementia care constraints, some review work mainly presented the summarise of assistive technologies in dementia care. 
Building upon these technological breakthroughs and in recognition of the limitations within traditional dementia care, several review studies have primarily focused on providing comprehensive summaries of assistive technologies in dementia care~\cite{pappada2021assistive,sriram2019informal,dada2021intelligent}. These innovative technologies have the potential to provide crucial support in the realm of dementia care. 
In addition to these studies, the development of support applications designed for informal caregivers can significantly enhance dementia care by offering valuable training and much-needed support ~\cite{bui2023ehealth,rahmawati2023dementia}. These applications empower caregivers with the knowledge and assistance required to navigate the challenges of dementia care effectively. Moreover, the evolution of interactive technologies customized for individuals living with dementia has emerged as an increasingly significant area of interest within HCI. 
% In addition to these review studies, the other researchers conducted the review studies about smart health technologies and mobile technologies designed to aid individuals living with dementia~\cite{dada2021intelligent,koo2019examining}. These advancements technologies can benefit for dementia care. 
Capitalizing on these advancements in dementia care, this paper contributes a scoping review that encompasses the latest research spanning from 2010 to July 2023, culled from authoritative literature sources like ACM and IEEE. 
% Compared to existing review papers, such as some review studies about smart health technologies and mobile technologies designed to aid individuals living with dementia~\cite{dada2021intelligent,koo2019examining}, our review paper presented the general aspects of advancement technologies in dementia care and
In comparison to existing review papers, such as those exploring smart health technologies and mobile solutions for individuals with dementia~\cite{dada2021intelligent,koo2019examining}, our review paper takes a broader perspective, delving into the overarching advancements in dementia care technologies. 
% Unlike existing review papers, which often narrow their scope to specific aspects of dementia care technology, our review takes a panoramic view, encapsulating the overarching advancements in this field.
Furthermore, the primary focus of this review lies in critical facets of dementia care and endeavors to enrich the experiences of PwD. 
In addition, through a meticulous mapping of existing literature in ACM and IEEE, our paper culminates in the categorization of six classifications of technologies for dementia care: Daily Life Monitoring, Daily Life Support, Social Interaction and Communication, Well-being Enhancement, Cognitive Support, and Caregiver Support. These classifications can guide us through the rich landscape of technological innovations poised to transform dementia care.
% These technologies can illuminate the potential of technology in bridging the gaps that exist in dementia care, and address the distinctive challenges posed by this condition, providing insights into how technology can be harnessed to improve the quality of care. 
% Moreover, our review paper can illuminate the potential of technology in bridging the gaps that exist in dementia care, and addressing the unique issues faced by this illness, as well as giving insights into how technology might be used to improve care quality. 
Moreover, our review paper illuminated the potential of technology in bridging the existing gaps in dementia care and addressing the unique challenges posed by this disease. It offered invaluable insights into how technology can be strategically harnessed to enhance care quality, enriching the lives of individuals affected by dementia.

\section{Background}
% Dementia care is currently difficult issue as it aforementioned, and it plays a central role in dementia treatment.
% As aforementioned, dementia is a multifaceted syndrome characterized by a decline in cognitive function, presents a pressing global healthcare challenge. 
As aforementioned, dementia is a complicated syndrome characterized by a decline in cognitive function that presents a major worldwide healthcare concern. In recent years, the issue of dementia care has broadened, placing growing demands on individuals, families, and healthcare systems, particularly in the context of our aging population. This section begins with a succinct introduction of dementia care, followed by a presentation of existing research and the up-to-date technologies employed in the field of dementia care.
% dementia care is currently a challenging issue in the healthcare, and it also plays a critical role in dementia therapy. The existing approaches in dementia care can enhance   

% This scoping review employs a systematic search and analysis approach to identify and categorize research articles, conference papers, and relevant literature at the intersection of Aging, Dementia, and Healthcare/Care within the HCI domain. The review includes studies from the past decade, focusing on design principles, user-centered methodologies, and technological interventions targeting older adults and individuals with dementia.

%People with dementia: What is dementia. How is life typically for people with dementia/life situation. Struggles related to dementia. 

%Careres of people with dementia. Informal. Formal. How it affects them/their struggles. 

%Generelly, how technology can benefit people's lives, make it easier in different manners. Technology for people with different types of illnesses and/or disabilities (assistive technology?) The role of HCI in developing technology that help people with different kinds of disabilities/illnesses, etc. 
% People with dementia is often old, and thereby may struggle with their physical body. 
% Main issue for poeple with dementia is their cognitive dissability. Other types of technolgies made for people with cognitive dissabilities. 

\subsection{Dementia Care}
\subsubsection{Dementia Types}
%People with dementia: What is dementia. How is life typically for people with dementia/life situation. Struggles related to dementia. 
% Dementia is a decline in mental ability that interferes with daily life. It is not a single disease but a term used to describe changes in memory, thinking, and reasoning. Dementia can be caused by various neurodegenerative diseases, including Alzheimer's disease, vascular dementia, frontotemporal dementia, and dementia with Lewy bodies.  
In general, dementia types can be divided into six major categories: mold cognitive impairment (MCI), Alzheimer's disease (AD), Parkinson's disease dementia (PDD), Lewy body dementia (LBD), frontotemporal dementia (FTD), and Creutzfeldt-Jakob disease (CJD)~\cite{ono2023mortality,manzine2022potential,magdy2022cognitive,mehraram2020weighted,gublood}. MCI and Alzheimer's disease are the most prevalent dementia types. MCI is a stage of cognitive impairment that is visible but not severe enough to be categorized as dementia~\cite{portet2006mild}. It can be considered a transitional phase between normal aging and dementia. 
Alzheimer's is the most well-known form of dementia, characterized by progressive memory loss and cognitive decline ~\cite{cummings2002alzheimer}. Many studies in our literature review research referenced these two types of dementia. Moreover, these classifications provide vital insights into the diverse landscape of dementia, aiding in accurate diagnosis and tailored care.

\subsubsection{Dementia Diagnosis}
The process of diagnosing dementia entails assessing cognitive decline and the impact of this decline on an individual's daily activities~\cite{arvanitakis2019diagnosis}. This evaluation is typically conducted by a clinician and involves a thorough history assessment and a comprehensive mental status examination. 
To initiate the cognitive assessment, brief cognitive impairment screening questionnaires can be employed~\cite{weintraub2022neuropsychological}. Further diagnostic insight can be gleaned through the use of neuropsychological testing \cite{jacova2007neuropsychological}. In addition to these cognitive assessments, physical examinations and brain neuroimaging serve as invaluable diagnostic tools in identifying the etiology of dementia, including conditions, such as stroke or structural alterations within the brain ~\cite{arvanitakis2019diagnosis}. However, it's important to acknowledge that achieving a highly accurate clinical dementia diagnosis can be a formidable challenge, especially during the early stages when relying solely on these approaches. In response to this challenge, recent advancements in AI have brought forth promising developments.  
% Artificial intelligence (AI) algorithms combined with conventional magnetic resonance imaging (MRI) have shown promise in improving the diagnostic accuracy of different types of dementia~\cite{merl2022truth}. 
Merl et al.~\cite{merl2022truth} demonstrate how AI algorithms, coupled with conventional magnetic resonance imaging (MRI), hold potential for significantly enhancing diagnostic accuracy across various dementia types.
% Furthermore, recent methods and techniques with potential to be utilized for multimodal dementia diagnosis include speech features, facial expression, movement patterns, sleep behavior, and mood changes~\cite{palliya2021advances}.
Furthermore, there is an emerging landscape of innovative methods and techniques that have the potential to be harnessed for multi-modal dementia diagnosis. Palliya et al.~\cite{palliya2021advances} presented the use of speech features, facial expressions, movement patterns, sleep behavior, and mood changes as promising avenues in this quest for improved diagnostic precision. 
% Based on these clinical dementia diagnosis combine with existing AI, wearable technology and multimodal behavior analysis, it is feasible to conduct the early-stage daignosis of dementia with promising result.
% By harnessing the synergy of clinical dementia diagnosis alongside the integration of existing AI capabilities, wearable technology, and the nuanced insights of multimodal behavior analysis, there emerges a promising pathway towards achieving early-stage dementia diagnosis with remarkable efficacy and potential.
Using clinical dementia diagnosis combined with existing artificial intelligence, wearable technology, and multi-modal behavior analysis, it is feasible to conduct an early-stage diagnosis of dementia with promising results. 

\subsubsection{Conventional Dementia Care} 
%Careres of people with dementia. Informal. Formal. How it affects them/their struggles. 
% The provision of dementia care represents an additional complex challenge within dementia research. Traditional dementia care aims to increase overall well-being and promote independence for as long as possible in order to delay placement in a long-term care institution~\cite{rappe2005influence}. PwD will get day-to-day care from health workers who will monitor their vital signs and any further behavioral or psychiatric issues that occur, as well as regular medical checks by a physician. Caregivers play a crucial role in improving dementia patients' quality of life throughout the illness course because they can track disease development and assist with medicine intake, fall prevention, activating the patient, and engaging in care-tasks~\cite{abdollahpour2018positive}. They also give emotional support or focus on stress reduction. Psycho-education, support groups, individual counseling, and respite care programs are among the formal caregiver interventions~\cite{chien2005effectiveness}. These therapies emphasize non-health advantages such as improved illness understanding, communication, and general dynamic between caregiver and patient, as well as improved patient and caregiver quality of life.
Dementia care represents a multifaceted challenge within the realm of dementia research. Traditional dementia care primarily focuses on two key objectives: enhancing overall well-being and promoting independence for as long as possible to delay the necessity of placing individuals in long-term care institutions ~\cite{rappe2005influence}. PwD receive day-to-day care from healthcare professionals who monitor their vital signs and address any behavioral or psychiatric issues that may arise. Additionally, regular medical check-ups are conducted by physicians to ensure comprehensive healthcare management.
Caregivers play an indispensable role in improving the quality of life of PwD throughout the course of the illness. They track the progression of the disease, assist with medication management, implement fall prevention strategies, engage patients in daily activities, and offer emotional support, with a focus on stress reduction~\cite{abdollahpour2018positive}. Formal caregiver interventions encompass a wide range of approaches, including psycho-education, support groups, individual counseling, and respite care programs~\cite{chien2005effectiveness}.
These interventions extend beyond healthcare considerations, emphasizing non-health-related benefits, such as enhanced understanding of the illness, improved communication, and a more positive dynamic between caregiver and patient. These efforts finally aim to enhance the overall quality of life for both the patient and the caregiver.

% However, the underlying mechanisms behind dementia are not well understood. PwD frequently experience changes in their symptoms, which typically provide clinicians with a picture of the patient's functioning on which to base their disease treatment strategy~\cite{livingston2020dementia}; however, standard biomarkers do not capture these swings. Medication efficacy and illness progression rates are difficult to predict for each particular patient.
However, the intricate mechanisms underlying dementia remain a subject of ongoing research, and our understanding in this field is far from complete. 
PwD frequently undergo fluctuations in their symptoms~\cite{livingston2020dementia} and conventional diagnostic tools often struggle to capture these dynamic changes. This presents a significant challenge, as it can complicate the provision of effective dementia care.
Moreover, the demands of day-to-day healthcare for individuals with dementia can place considerable stress on caregivers, both emotionally and financially, making it a burdensome undertaking for many families. As a result, some early-stage people living with dementia may opt for in-home healthcare as an alternative. However, limited experience in dementia care can significantly compound the challenges faced by informal caregivers. 
Due to these challenges, a range of technologies, including assistive devices, telehealth and remote monitoring, robotics, sensory and environmental enhancements, among others, have emerged as promising solutions to provide valuable support for dementia care.
\subsection{The Advancements of Technologies in Dementia Care}
% In general, PwD can receive appropriate care in clinical settings but not in home environment. The main reason is the informal caregivers with limit experience and usually these people are their family members, even they live alone. Improving the well-being for PwD and training dementia caregivers has become current challenges. 
In general, PwD often receive appropriate care in clinical settings; however, providing the same level of care in a home environment poses unique challenges. Informal caregivers, who are frequently family members, even if the PwD lives alone, may have limited experience in dementia care. This presents a pressing need to enhance the well-being of PwD and provide comprehensive training for dementia caregivers.
% In order to solve these issues, the existing technologies can offer alternatives and these technologies have made significant contributions to dementia care, improving the quality of life for both patients and caregivers. These advancements have evolved in various ways, including diagnosis, treatment, monitoring, and support. In these approaches, monitoring and support plays an central in dementia care.
To address these challenges, existing technologies have provided valuable alternatives, making substantial contributions to dementia care and enhancing the quality of life for both patients and caregivers. These advancements have taken shape in various dimensions, spanning diagnosis, treatment, monitoring, and support. Particularly, monitoring and support hold a pivotal role in the landscape of dementia care. 
% In monitoring context, Unobtrusive monitoring (UM) technologies are being explored to support extended independent living of people with dementia (PwD) at home~\cite{wrede2022create}. These technologies aim to remotely monitor the lifestyle, health, and safety of PwD, providing gains such as objective surveillance and timely interventions, while also addressing concerns such as information overload and less human interaction~\cite{mclaughlin2022evaluating}.
As mentioned in a recent research by Wrede et al.~\cite{wrede2022create}, there is a rising investigation into Unobtrusive Monitoring (UM) technology to allow the extended independent life of PwD in their homes. These technologies are intended to remotely monitor the lifestyle, health, and safety of PwD, providing advantages such as objective surveillance and prompt interventions. However, they also raise considerations such as potential information overload and reduced human interaction~\cite{mclaughlin2022evaluating}.
Moreover, several support technologies, such as assistive technologies for PwD and their caregivers~\cite{ienca2018ethical,sriram2020carers}, can enhance their quality of life as well.

% With the development of AI, Large Language Model (LLM), Wearable Technologies, Robotics, etc., more potential technologies can be implemented in existing technologies in dementia care and these innovative technologies can extremely improve the well-being for PwD and their caregivers. At the same time, these technologies can also be reduce the burden from families and hospitals. 
% These advantages motivate us to conduct our literature review studies and simultaneously explore the future research in dementia care. 
As Artificial Intelligence (AI), Large Language Models (LLMs), Wearable Technologies, Robotics, and other cutting-edge innovations continue to advance, they offer substantial potential for integration into existing dementia care technologies. These innovative solutions have the power to significantly enhance the well-being of PwD and their caregivers. Simultaneously, they can alleviate the burden on both families and healthcare institutions. The advantages of these technologies serve as a compelling motivation for us to conduct comprehensive literature review studies and embark on explorations into the future of research in dementia care. Their transformative impact on the field holds great promise for improving the lives of PwD and their caregivers while optimizing the overall care ecosystem.
\section{Method}
% The paper categorizes research findings into several key themes, including:
% a. Assistive Technologies: Exploration of interactive tools, wearable devices, and smart environments to enhance the daily activities and independent living of older adults and individuals with dementia.
% b. Cognitive Support: Investigation of technology-based interventions to support cognitive function, memory recall, and decision-making processes in dementia care.
% c. Social Interaction and Communication: Analysis of digital platforms and interfaces that facilitate social engagement, communication, and emotional well-being for older adults, particularly those experiencing social isolation.
% d. Usability and Accessibility: Evaluation of design principles, user interfaces, and accessibility features that ensure the usability and acceptance of technology among older adults and individuals with cognitive impairments.
% The goal of our literature review is presenting the current trend about technologies in dementia care, and also present some hints for the future research. It means that the collected research from recent years can contributes to the future research in dementia care. By integrating the wide variety of literature about dementia care, we aims to provide a meaningful way of understand the existing technologies in dementia care. This section presents research questions firstly, and then describe the publication search process according to these research questions.

Our literature review aims to accomplish two primary objectives: firstly, to present the current trends in technologies for dementia care, and secondly, to offer insights and directions for future research in this field. Our intention is to bridge the gap between recent research findings and their potential contributions to the future of dementia care. Through a comprehensive analysis of diverse literature sources on dementia care, we aspire to provide a meaningful framework for understanding the existing landscape of technologies in this domain.
In this section, we have formulated two research questions that will guide our review process. These research questions serve as the foundation upon which we will systematically conduct our literature search, ensuring that our exploration aligns with the key aspects of technology in dementia care. Subsequently, we will outline our approach to the publication search process, illustrating how it corresponds to these research questions.
\subsection{Research Question}
% In order to explore the research status of how HCI techniques are employed in the field of dementia care, we formulated a set of pivotal research questions that serve as the foundation for this review.
% As aforementioned, dementia care strives to enhance the lives of individuals with dementia and the well-being of their caregivers. 
% % Based on these goals, we presented the following two research questions concerning dementia care.
% With these goals in mind, we have formulated the following two research questions to guide our exploration of dementia care:
As aforementioned, dementia care strives to enhance the lives of individuals with dementia and the well-being of their caregivers. 
In order to explore the research status of how ongoing technologies are employed in the field of dementia care, we have formulated the following two research questions to guide our exploration of dementia care:

\begin{itemize}
    \item \textbf{RQ1:} Which technology-based dementia care strategies can effectively improve the life quality of people living with dementia?
    \item \textbf{RQ2:} Which technology-based stress-coping approaches can significantly alleviate caregivers' strain and distress for people living with dementia? 
\end{itemize}

% These two research questions prompted us to conduct a literature review to find answers. Currently, technology-based dementia care, in addition to traditional dementia care approaches, can proficiently improve the living quality of dementia patients combined with existing AI, wearable technologies, etc. 
At present, the fusion of technology-based dementia care with traditional approaches has the potential to significantly enhance the quality of life for PwD. This synergy, complemented by emerging technologies, such as Artificial Intelligence (AI), wearable devices and robotic, holds promise in revolutionizing dementia care and support. 
% Based on the first research question, we can trigger various key words for use in literature searches, such as "technology-based", " dementia care", "dementia patients," "life quality," etc. 
Addressing the first research question necessitates the exploration of pertinent keywords in our literature searches, including terms, such as "technology-based", "dementia care", "older people with dementia", and "quality of life". By leveraging these keywords, we aim to unearth a wealth of knowledge that can inform our quest to optimize the well-being of individuals living with dementia.
% Furthermore, if we can provide some assistance to dementia caregivers and reduce their workload or stress, persons with dementia will be able to get appropriate care while keeping their emotional well-being. As a consequence, another way to improve the quality of life for people with dementia is to provide some technology-based assistance and help dementia carers with these tools. From the second research questions, we can also get some relevant key words, like "dementia caregivers", "technology-based", etc.  
Moreover, our exploration of current HCI research for dementia care extends to technologies for providing essential assistance to their dedicated caregivers. These caregivers often bear a significant burden of responsibility, which can lead to stress and emotional strain. To address this, integrating technology-based solutions into their care routines can offer much-needed support. Thus, another avenue to improve the lives of individuals with dementia is by implementing technology-based tools that assist and empower dementia caregivers.
For the second research question, we will explore relevant keywords, such as "dementia caregivers" and "technology-based" to unearth valuable insights into innovative ways of supporting and bolstering the invaluable work of these caregivers.

\subsection{Search Strategy}
\label{search_strategy}
% Having identified possible keywords from these two research question, we also used patients, intervention, comparison, outcome(PICO)~\cite{eriksen2018impact} as our literature search strategy tool. 
% We utilized patients, intervention, comparison, outcome (PICO)~\cite{eriksen2018impact} as our literature search strategy tool after identifying potential keywords from these two research questions. In this search strategy tool, "P" can be also defined as person, population or 
% patients, and it aims to find a specific population we need focus on. In our literature review, we mainly focus on patients with dementia and their caregivers. Intervention in our literature review involve in advancement technologies, such as assistive technologies, wearable technologies, robotic technologies, etc. In our context, it includes dementia care in home setting or hospital environments. The outcome of dementia care is to improve the quality life for PwD and their caregivers. 
% In our literature search strategy, we employed the Patients, Intervention, Comparison, and Outcome (PICO) framework, as outlined by Eriksen et al.~\cite{eriksen2018impact}. Within this systematic approach, the "P" facet encompasses the specific population of interest, which, in our case, revolves around individuals with dementia and their dedicated caregivers.
In our literature search strategy, we embraced the comprehensive Patients, Intervention, Comparison, and Outcome (PICO) framework, drawing inspiration from the guidelines set forth by Eriksen et al.~\cite{eriksen2018impact}. Within this systematic approach, the "P" facet of our approach intricately focuses on individuals with dementia and their dedicated caregivers, reflecting our commitment to a patient-centered perspective.
The "I" component of our strategy encompasses a broad range of technological interventions, encompassing assistive technologies, wearable devices, robotic advancements, and related innovations. These interventions are examined within the context of dementia care, both in home settings and hospital environments.
% The "C" in PICO method means comparison or control group, however there is no comparing or control group in our case and we used "context" instead of.
While the traditional "C" in PICO typically refers to a Comparison or Control group, we tailored this element to suit our distinct case. Instead of a direct comparison, we spotlight the significance of "Context." This adaptation allows us to explore and understand the broader environmental and situational factors that shape the effectiveness of interventions in our chosen field.
Our focus on the "O" or Outcome aspect centers on the overarching goal of dementia care, which is to enhance the quality of life for Persons with Dementia (PwD) and alleviate the burdens faced by their caregivers. This emphasis underscores the ultimate objective of our literature review: to identify technologies and strategies that contribute positively to the well-being of both PwD and their dedicated caregivers.
It is not difficult to confirm our keywords and conduct literature search in dementia care using this structured PICO method.

\begin{enumerate}
    \item \textbf{P} - old people with dementia \textbf{OR} elderly live with dementia \textbf{OR} people with aging and dementia \textbf{OR} older individuals with dementia \textbf{OR} senior with dementia \textbf{OR} dementia carer \textbf{OR} dementia caregiver 

    \item \textbf{I} - HCI \textbf{OR} assistive technology \textbf{OR} welfare technology 
    
    \item \textbf{C} - dementia care \textbf{OR} dementia nursing care 

    \item \textbf{O} - quality of life \textbf{OR} well-being 
\end{enumerate}

% We aim to provide an up-to-date analysis of recent advancements in technological HCI within the context of dementia care, focusing on research published in the last 13 years (2010--2013). To achieve this, we employed scoping reviews to gain insights into the evolving landscape of research in this domain. Scoping reviews serve as a valuable tool for summarizing research findings, mapping the extent and nature of emerging research areas, and pinpointing potential research opportunities~\cite{tricco2016scoping}. In our data collection and selection process, we adhered to the Preferred Reporting Items for Systematic Reviews (PRISMA) guidelines~\cite{selccuk2019guide} and Meta-Analyses Extension for Scoping Reviews (PRISMA-ScR)~\cite{page2017evaluations} framework, commonly utilized for crafting review papers. As it shown in Figure~\ref{fig:search_flowchart}, it provides an illustration of our literature selection process, comprising four iteration stages: data collection, initial filtering, research question exclusion, and full-text screening. To initiate our study, we identified relevant search terms based on prior knowledge and executed a systematic query across two prominent databases (n=2) renowned for publishing a vast repository of HCI academic papers: ACM Digital Library and IEEE Xplore.
\subsection{Literature Selection}
\begin{figure*}[h]
\centering
  \includegraphics[width=0.6\linewidth]{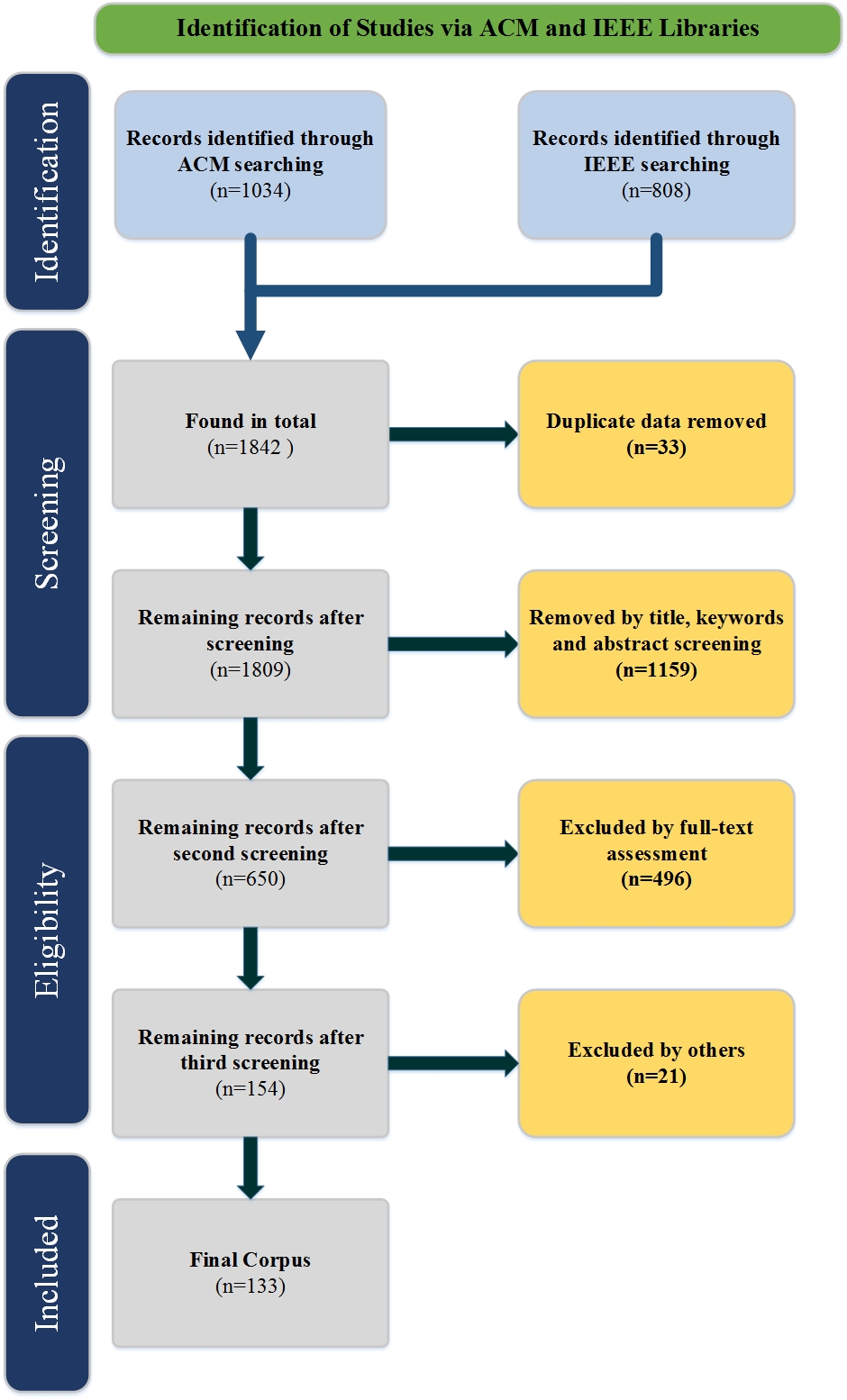}
  \caption{The PRISMA flowchart of selection and refinement process in our literature review}
  \label{fig:search_flowchart}
\end{figure*}

Our aim is to offer an up-to-date and comprehensive analysis of recent advancements in the realm of technological HCI within the context of dementia care, with a particular focus on research published within the last 13 years (2010-2023). To achieve this objective, we have harnessed the power of scoping reviews, a valuable approach that enables us to synthesize research findings, map the scope and nature of emerging research areas, and identify potential avenues for future investigation~\cite{tricco2016scoping}.
Throughout our literature collection and selection process, we adhered to the Preferred Reporting Items for Systematic Reviews (PRISMA) guidelines~\cite{selccuk2019guide} and the Meta-Analyses Extension for Scoping Reviews (PRISMA-ScR) framework~\cite{page2017evaluations}, widely recognized methodologies for constructing comprehensive review papers.
As illustrated in Figure~\ref{fig:search_flowchart}, our literature selection process unfolded across four iterative stages: data collection, initial filtering, research question exclusion, and full-text screening. 
To commence our study, we identified pertinent search terms based on our existing knowledge and subsequently executed a systematic query across two esteemed databases — ACM Digital Library and IEEE Xplore, both renowned for hosting a vast repository of HCI academic papers.

% \begin{enumerate}
%     \item In the initial query, we gathered a substantial set of 1842 papers (n=1842).
%     \item A filtering process was subsequently applied based on evaluating titles, keywords, abstracts, and full-text content. This operation led to the retention of 656 papers.
%     \item The remaining records underwent a further exclusion phase, during which papers not aligned with our research objectives were removed. Then, we curated a collection of 160 papers.
%     \item Finally, we conducted a manual screening phase, identifying relevant papers by full-text and excluding review articles. As a result, our final corpus consists of 139 papers (n=139), representing a substantial collection compared to prior review studies.
% \end{enumerate}

\paragraph{Identification} 
% In identification process, we used the keywords as aforementioned and then put these keywords into the ACM and IEEE libraries, as well as specified the year from 2010 to 2023. These two literature database well related with our research and most HCI conference or journals are collected into these two literature database. We obtain the required literature from ACM (1034) and IEEE (808) in these two libraries using the keywords which was mentioned in \ref{search_strategy} and ranged the year from 2010 to 2023 in the full text.
During the identification process, we utilized the previously mentioned keywords. These keywords were input into both the ACM and IEEE libraries, renowned repositories encompassing a comprehensive collection of literature closely aligned with our research objectives. Specifically focusing on the years 2010 to 2023 ensured that our search yielded the most current and relevant publications. Our search keywords, detailed in \ref{search_strategy}, efficiently navigated the vast repositories, resulting in the acquisition of pertinent literature. From the ACM database, we retrieved 1034 publications, while the IEEE database contributed 808 publications. Finally, we found 1842 relevant pieces of literature.

\paragraph{Screening}
To ascertain the relevance of the identified sources to our research objectives, a meticulous manual screening procedure was implemented by two of the authors. This involved a thorough examination of titles, keywords, and abstracts right from the outset. Prior to the screening, 33 duplicate pieces of literature and 12 review articles were removed, streamlining the subsequent evaluation process. 
% Furthermore, if certain papers were too short or in poster style, they were removed. However, some papers, such as those from CHI's late break work, were included since they presented brief stories relevant to our work and were formatted as short papers rather than posters.         
During the screening procedure, we encountered papers associated with dementia care that prompted thoughtful consideration of their alignment with our research topics. Additionally, we identified innovative technologies within this context, sparking interest for potential application in dementia care. 
% The inclusion/exclusion criteria were further refined to encompass papers involving individuals with aging and dementia. 
The inclusion/exclusion criteria have been expanded to include whether or not the papers cover elderly people living with dementia.
After applying these criteria, a total of 1159 pieces of literature were excluded during the screening process, resulting in a refined collection of 650 records that remained pertinent to our research objectives.

\paragraph{Eligibility}
To ensure that the alignment of our final selection with the eligibility criteria, the same two authors revisited these papers and exercised discernment. In this refining process, we discerned papers that, while mentioning dementia care, primarily delved into research directions unrelated to our specific objectives, for example by mainly focusing on technical development processes. These were promptly excluded from our consideration. Moreover, a comprehensive examination of the full-text papers was undertaken to ascertain their relevance in addressing either \textbf{RQ 1} or \textbf{RQ 2}. Additionally, we evaluated whether the novel technology or design presented in the paper could potentially benefit PwD or dementia caregivers. Furthermore, special consideration was given to the nature of certain publications that did not align with the definition of full research articles, such as workshops, keynote addresses, posters, etc. However, some papers, such as those from CHI's late break work, were included since they presented brief stories relevant to our work and were formatted as short papers rather than posters. Adhering to these criteria, a total of 496 pieces of literature were excluded during the eligibility assessment, resulting in a streamlined collection of 154 records. 

\paragraph{Final Corpus}
% After determining eligibility, the remaining articles total 154, and these papers will be divided into two parts depending on different databases. Each author read the allocated papers in their entirety and examined them using the previously stated criteria. During this phase, certain papers were difficult to decide whether to exclude or include based on the preceding criteria, therefore they were marked and the authors decided in the final group discussion. 
% The culminating refined corpus now comprises 133 papers that precisely meet our criteria and research objectives.
Following the eligibility determination, the remaining articles total 154, and these papers will be divided into two sets depending on different literature databases. Each author then undertook the comprehensive reading and examination of the allocated papers, applying the predefined criteria.
Throughout this phase, the authors met regularly to compare and discuss their findings, especially focusing on papers posing difficulties in terms of inclusion or exclusion based on the established criteria. Recognizing the complexity of these decisions, such papers were appropriately marked for subsequent deliberation during the final group discussion. This collaborative decision-making process ensured a thorough and nuanced evaluation. In culmination, the refined corpus now stands at 133 papers, a meticulously curated selection precisely aligned with our criteria and overarching research objectives.

\subsection{Data Extraction}

The process of data extraction from our final corpus marks a crucial stride in addressing our research inquiries. We systematically gathered standard information such as authors, affiliations, titles, abstracts, publication type, and publication year from the included publications. Moreover, delving deeper into each paper allowed us to capture more nuanced details, including study objectives, motivation, technological advancements, methodologies, and environmental settings.
Our focus remained sharp on synthesizing major insights from each study, coupled with identifying the technologies employed, with a keen eye on addressing either \textbf{RQ 1}, \textbf{RQ 2}, or both. Beyond the technological realm, our analysis extended to exploring environmental settings and caregiving techniques such as health monitoring, robotic assistance, cognitive support, wearable technology, and more.
Furthermore, the extraction process encompassed crucial details related to study design and participant information from the 133 papers. The culmination of these summaries and extracted critical information positions us for a comprehensive final analysis. This analysis is poised to unveil existing dementia care technologies while concurrently informing the exploration of potential technologies for future dementia care studies.
Moreover, the final summation and analysis derived from each chosen literature source hold significant promise in validating various categorizations and deepening our understanding of the primary research areas in dementia care technologies. Additionally, establishing clear categorizations and conducting thorough analysis can serve as a robust framework for exploring future avenues of research in dementia care.

\section{Results}
% In this section, we mainly present on the results of existing technologies in dementia care during the recent years. According to these current technologies, we summarise these research topics and divided them into 6 distinct categories, including daily life monitoring, daily life support, social interaction and communication, well-being enhancement, cognitive Support and caregiver support.
In this section, our primary focus is to present the results of our analysis, with an emphasis on recent technological advancements in the domain of dementia care. Grounded in these contemporary technologies, we have systematically categorized our findings into six distinct and coherent themes. These thematic categories includes \emph{Daily Life Monitoring}, \emph{Daily Life Support}, \emph{Social Interaction and Communication}, \emph{Well-being Enhancement}, \emph{Cognitive Support} and \emph{Caregiver Support}.
By categorizing our findings into these six research themes, we aim to offer a structured and insightful panorama of the current technological milieu within dementia care. 

\subsection{The Result in Literature Searching}
\begin{figure*}[htbp]
\centering
  \includegraphics[width=\linewidth]{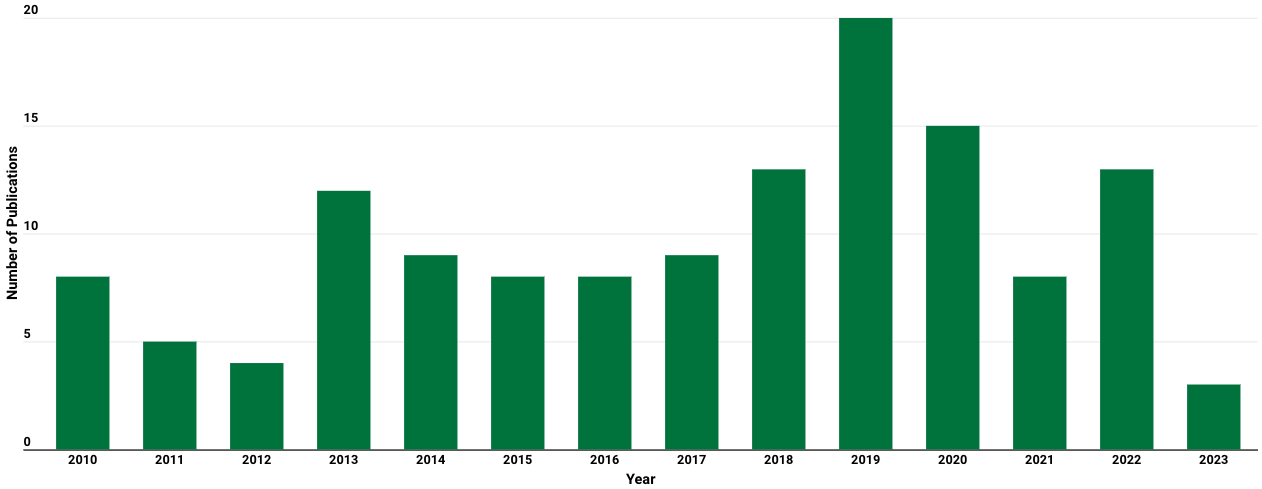}
  \caption{The research agenda regarding current technologies in dementia care during the years of 2010--2023.}
  \label{fig:trend_year}
\end{figure*}

Our research review was carried out within the comprehensive ACM and IEEE libraries, ultimately leading us to the selection of 133 highly relevant publications for our literature review.
To visually represent our findings and the dynamic landscape of dementia care technology, we have utilized histograms, as exemplified in Figure~\ref{fig:trend_year}. This graphical representation clearly illustrates a discernible trend: a notable increase in new dementia care technologies, particularly in the year 2019. 
Many assistive and interactive technologies are utilized to help with dementia care and enhance PwD well-being, such as the usage of VR to give accessible experience for persons with dementia~\cite{tabbaa2019bring}.
This surge in research and innovation aligns with the growing concern surrounding dementia disorder, propelled by the challenges posed by an aging population. It is also important to note that the research review was carried out during mid 2023, so the literature from 2023 only represents papers published between January and July.
% With the aging demographic, dementia has emerged as a prominent and pressing concern. 
Moreover, formal or professional dementia care alone is unable to fully address the multifaceted needs of PwD. 
% The primary living environment for dementia sufferers is often their own home, where they may live alone, or their family members may have limited expertise in dementia care. 
The primary living environment for dementia patients tends to be their own home, where they may live alone or with family members who have limited knowledge in dementia care.
This indicates the significance of assistive and interactive technologies that can provide meaningful support and assistance. 
% As a result, innovative research in dementia care technology in recent years is becoming increasingly indispensable in our quest to improve the lives of those affected by this condition in recent years.
As a consequence, recent years have seen a growing indispensability of innovative research in dementia care technology, furthering our quest to enhance the lives of those impacted by this condition.

% Another figure depicts current year trends in AMC and IEEE on present technology in dementia care. From the Figure~\ref{fig:trend_two_library}, it also shows that there are more innovative technologies on dementia care in recent years. Moreover, there are more relevant publications in ACM than in IEEE. In fact, it presents sensor-based technology which may be employed in dementia care in IEEE. Yet, these research only proposed the prototype of innovative technology and it is hard to say these technologies can be implemented in real dementia care environment. Furthermore, there are more research in IEEE focused on dementia detection using multi-sensors combined with deep learning method. In ACM library, many interactive technologies are proposed and these technologies can enhance PwD experience and well-being. 

\begin{figure*}[h]
\centering
  \includegraphics[width=\linewidth]{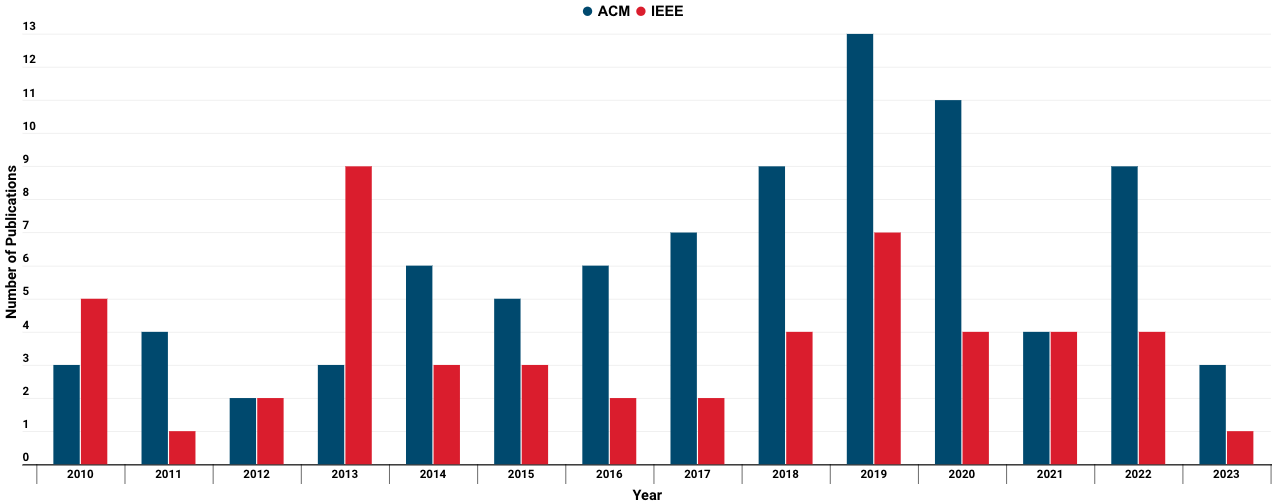}
  \caption{The research agenda in current technologies for dementia care in ACM and IEEE during the years of 2010 -- 2023.}
  \label{fig:trend_two_library}
\end{figure*}

Another figure illustrates the current trends in ACM and IEEE regarding technology in dementia care. As depicted in Figure~\ref{fig:trend_two_library}, it becomes evident that there is a notable upswing in innovative technologies for dementia care in recent years. Furthermore, it is worth noting that ACM has a higher volume of relevant publications in this domain compared to IEEE.
A closer examination reveals that IEEE publications often revolve around sensor-based technologies that have the potential for application in dementia care. However, it is important to acknowledge that many of these research efforts are in the prototype stage, and their practical implementation in real-world dementia care environments remains uncertain.
Conversely, ACM's research landscape appears to focus more on assistive and interactive technologies designed to enhance the experiences and well-being of PwD. These technologies hold promise in positively impacting the lives of individuals living with dementia, improving their overall quality of life.
It is also noteworthy that IEEE research tends to delve into dementia detection using multi-sensor setups coupled with deep learning methods, thus indicating an emphasis on diagnostic and monitoring aspects. In contrast, ACM's focus on interactive and assistive technologies underscores the potential for enhancing the day-to-day lives of those affected by dementia.
Together, these trends in ACM and IEEE publications underscore the multifaceted nature of research in dementia care technology, with a wide array of approaches and potential applications being explored.

\subsection{The Categories of Technology in Dementia Care}
% From our selected 133 paper, it consists of various technologies in dementia care, including wearable device, mobile application, robotic technology, sensors-based technology, and among others. It is not easy to define their categories and it is possible to group these technologies based on their application. For instance, sensor-based technology for daily activities and sleeping behaviors monitoring can be grouped as daily life monitoring. In some mobile applications, reminder function and navigation support can be grouped as daily life support. Moreover, most of robotic technologies, such as social robots, voicebot, etc, can assist with social interaction and communication fro PwD. Some interactive technologies can enhance PwD's well-being and these technologies can be called well-being enhancement. The other application can be utilised to train the PwD cognitive or provide some support for PwD's cognitive. These cased can be grouped into cognitive support. The remaining about support application for caregivers can be grouped as caregiver support. 
Within our selection of 133 papers, we encounter a rich landscape of technologies aimed at improving dementia care. These technologies encompass a wide spectrum, ranging from wearable devices and mobile applications to robotic innovations and sensor-based systems. While categorizing them into rigid classifications can be challenging, a more nuanced approach emerges when we group these technologies based on their intended applications and functions.
For instance, technologies rooted in sensor-based monitoring, which track daily activities and sleeping behaviors, can be effectively grouped under the overarching category of "daily life monitoring". This category encapsulates solutions that diligently observe and assess the day-to-day routines and sleep patterns of individuals with dementia. Similarly, mobile applications that feature reminder functions and navigation support naturally fall within the purview of "daily life support". These applications are designed to provide practical assistance to individuals with dementia, offering guidance and support to navigate their daily lives effectively. Robotic technologies, including social robots and voice-assistive bots, shine as beacons of "social interaction and communication" for PwD. They serve as invaluable tools for fostering connections and facilitating meaningful engagement, addressing a fundamental aspect of dementia care. Moreover, certain interactive technologies are strategically focused on "well-being enhancement" among PwD. These encompass a wide range of interventions, from immersive virtual reality experiences to personalized well-being programs, all aimed at raising the overall quality of life for those living with dementia. In addition, technologies tailored to "cognitive support" are designed to provide cognitive assistance and support to PwD. These solutions address cognitive challenges and offer valuable support to enhance their mental faculties. Finally, an important facet of our categorized technologies pertains to "caregiver support". These applications and tools are geared toward supporting caregivers of individuals with dementia, providing them with resources, training, and assistance in their caring roles. By adopting this nuanced categorization based on the intended applications, we aim to offer a more comprehensive understanding of these diverse technologies and their potential impact within the context of dementia care.

% The achieved categories of technologies in dementia care finally divided into 6 different classification. Based on these 6 categories, we made a histogram using the selected paper and we found most of research about technologies in dementia care are mainly focus on  daily life monitoring, social interaction and communications, well-being enhancement, as it shown in Figure~\ref{fig:categories_technology}. Although there are not too much paper about the other three categories, it doesn't mean that these three categories technologies are not important for PwD and their caregivers. In the other words, current technologies may not afford enough perfect to support for PwD and caregivers. With the advancement of AI and sensor technologies, more research will also focus on these three categories in the future. 

\begin{figure*}[htbp]
\centering
  \includegraphics[width=\linewidth]{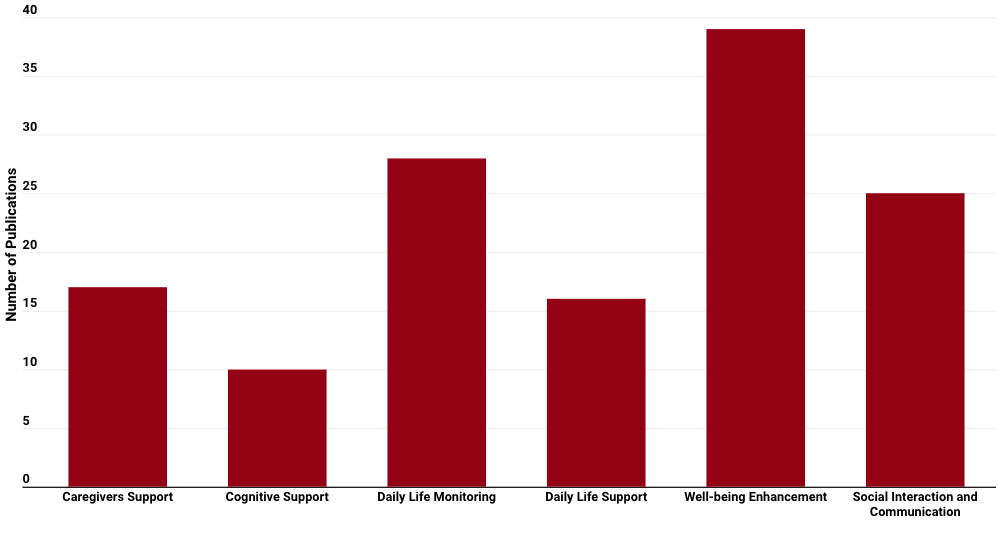}
  \caption{The number of publications regarding current technology categories in dementia care.}
  \label{fig:categories_technology}
\end{figure*}

% The technologies in dementia care have been effectively categorized into six distinct classifications. 
When we examined the distribution of research across these six distinct categories by utilizing the selected papers, a clear trend emerged. That is, as Figure~\ref{fig:categories_technology} illustrates, a significant portion of the research in dementia care technologies predominantly concentrates on three primary areas: daily life monitoring, social interaction and communication, and well-being enhancement.
However, it is important to emphasize that the relatively lower number of papers in the remaining three categories—cognitive support, caregiver support, and daily life support—should not diminish their significance. These categories are equally vital for the well-being of PwD and their caregivers. The current emphasis on these areas may reflect the evolving nature of technology adoption in dementia care.
In essence, it is crucial to recognize that existing technologies may not yet provide a comprehensive solution to address all the multifaceted needs of PwD and their caregivers. Nevertheless, with the rapid advancements in Artificial Intelligence (AI), interactive technologies and sensor technologies, we can anticipate a growing body of research in these underrepresented categories in the future. These advancements hold the potential to further enrich and diversify the technological landscape in dementia care, offering innovative solutions to enhance the lives of those affected by this condition.

% The histogram from Figure~\ref{fig:trend_categories_technology} shows the distribution of the categories of existing technologies in dementia care during the recent years. As it depicted in Fiure~\ref{fig:trend_categories_technology}, "well-being enhancement" and "Social Interaction and Communication" are ongoing hot research concerning technologies in dementia care. "Daily Life Monitoring" presents declining trend in recent 5 years and "Daily Life Support" presents the opposite trend from this figure. The main reason is that we only use literature database of ACM and IEEE libraries. The publications about these research in the other database are not presented here. 
% With the advancement of interactive technologies and robotics, more research and application can provide some support for PwD and their caregivers' daily life, PwD' skills of social interaction and communication, and enhance their well-being after experiencing these technologies. 

\begin{figure*}[htbp]
\centering
  \includegraphics[width=\linewidth]{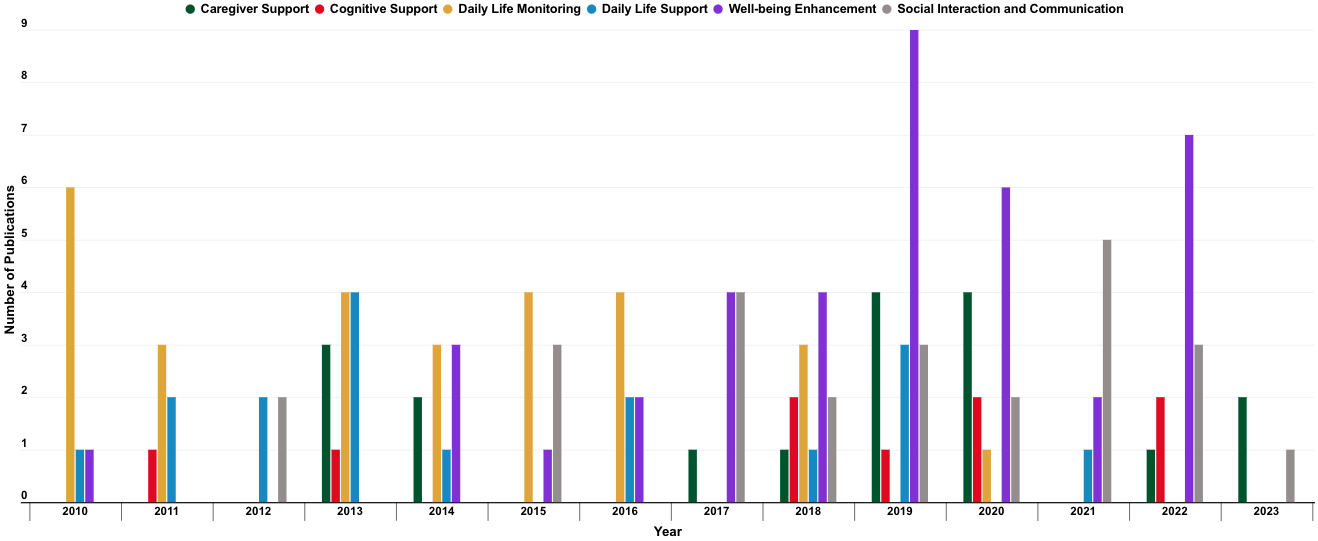}
  \caption{The research agenda of individual category of current technology in dementia care during the years of 2010 -- 2023.}
  \label{fig:trend_categories_technology}
\end{figure*}

The histogram depicted in Figure~\ref{fig:trend_categories_technology} offers a visual representation of the evolving landscape of technology categories in dementia care in recent years. Notably, "well-being enhancement" and "Social Interaction and Communication" emerge as prominent and consistently researched areas within the realm of dementia care technologies. Interestingly, the data also reveals contrasting trends in "Daily Life Monitoring" and "Daily Life Support". "Daily Life Monitoring" demonstrates a declining trend over the past five years, while "Daily Life Support" exhibits the opposite trajectory. It is essential to note that these trends are influenced by the specific literature databases we utilized, namely ACM and IEEE libraries. Publications from other databases may present a different perspective on these areas.
The observed trends underscore the dynamic nature of research in dementia care technology. The rise of interactive technologies and robotics has paved the way for innovative research and applications that directly impact the daily lives of PwD and their caregivers. These technologies hold the potential to enhance various facets of dementia care, ranging from supporting daily life routines to improving social interactions, and ultimately, enhancing the overall well-being of those who rely on them.
As technology continues to advance, we anticipate that these trends will evolve further, with a growing emphasis on providing comprehensive support for PwD and their caregivers across all aspects of dementia care.

\subsection{Review Topics of Current Technologies in Dementia Care}
% From the Table~\ref{tab:catergories_dementia_care}, it presents the selected relevant paper and their corresponding categories
Table~\ref{tab:catergories_dementia_care} provides an overview of the selected relevant papers and their corresponding categories.
Through the thematic analysis of the data material, as previously discussed, we have identified and categorized technology into six distinct areas aimed at enhancing the lives of individuals with dementia: 1) technologies for \emph{Daily Life Monitoring}, 2) technologies for \emph{Daily Life Support}, 3) technologies for \emph{Social Interaction and Communication}, 4) technologies for \emph{Well-being Enhancement}, 5) technologies for \emph{Cognitive Support}, and 6) technologies for \emph{Caregiver Support}. In this section, we take a closer look at each of these categories. 
% Based on the thematic analysis of the data material as it mentioned before, we found six distinct categories of technology developed to improve the lives of people with dementia.

\begin{table}[htbp]
    \centering
    \caption{The distribution of technology categorization to dementia care in the corpus}
    \label{tab:catergories_dementia_care}
    \begin{tabular}{ccc}
    \hline
    \makecell[c]{The Technologies-based \\ Approaches to Dementia Care} & Reference & Number \\
    \hline
    \textbf{Daily Life Monitoring} & & \textbf{28}\\
    \hdashline
    Mobile-based Monitoring & \cite{mulvenna2010designing, sposaro2010iwander, solanas2013m, shree2014design,geddes2010cloud, vuong2011feasibility} & 6 \\
    Sensor-based Monitoring & \cite{biswas2010sensor,wang2010monitoring, schikhof2010will, wan2014addressing,  stavropoulos2016multimodal, radziszewski2016ambient, wang2016location, nesbitt2018reliability, paletta2018playful,wan2016design} & 11 \\
    Sleeping Monitoring & \cite{nikamalfard2011sleep, ehleringer2013wearable, gong2015home} & 3 \\
    Application Monitoring & \cite{osman2013utilising, tupakula2014secure, guldenpfennig2015proxycare, lam2015smartmind, enshaeifar2020digital, aljehani2018icare} & 6 \\
    Wandering Management & \cite{muller2013dealing, wan2015managing} & 2 \\
    \hline
    \textbf{Daily Life Support} & & \textbf{16} \\
    \hdashline
    Mobile Application & \cite{armstrong2010developing, camarena2012weremember, yamagata2013mobile, zmily2013alzheimer} & 4 \\
    Smart Home & \cite{ikeda2011remote,hwang2012using, amiribesheli2016towards,thapliyal2017smart,grave2021requirement} & 5 \\
    Assistive Robot& \cite{begum2013performance, hebesberger2016lessons, di2019robotic} & 3 \\
    Assistive Technologies& \cite{favilla2013touch, peters2014automatic,zanwar2018assistive,jonsson2019reminder} & 4\\
    \hline
    \textbf{Social Interaction and Communication} & & \textbf{25} \\
    \hdashline
    Robotic Technology & \makecell[c]{ \cite{shibata2012therapeutic, rudzicz2015speech, sandoval2017co, perugia2017modelling, abdollahi2017pilot,joshi2019robots,iwabuchi2019communication,cruz2020social,itai2020development,zhou2021assistant},\\ \cite{taylor2021exploring,zubatiy2021empowering,striegl2021designing,abdollahi2022artificial,yuan2023cognitive}  } 
    & 15 \\
    Assistive Technology & \cite{green2012assisting, schelle2015tactile, goto2015remote, chinaei2017identifying, nakatani2018generating, unbehaun2018exploring,unbehaun2018facilitating, munoz2021evaluating} & 8 \\
    Mobile Application & \cite{thoolen2022livingmoments, munoz2022evaluating} & 2 \\
    \hline
    \textbf{Well-being Enhancement} & & \textbf{39} \\
    \hdashline
    Reminiscence Therapy&\cite{iwamoto2014study,gundogdu2017activating,filoteo2018evaluation,huber2019tangible,edmeads2019designing,ferreira2019musiquence,sun2022web} & 7\\
       \hdashline[0.5pt/5pt]
    \textit{UX Enhancement} &  & \\
    ~~Interactive Technology & \makecell[c]{\cite{siriaraya2014recreating,bennett2016rekindling,lazar2016designing,seymour2017ami,eisapour2018participatory,lopez2018acceptance,tabbaa2019bring,feng2019livenature,foley2019printer,mertl2019traumreise},\\ \cite{unbehaun2020social,sas2020supporting,houben2020turnaround,thoolen2020ambientecho,karaosmanoglu2021lessons,houben2022enriching,marchetti2022pet,matsangidou2022free,czech2022just} }& 19 \\
    ~~Everyday Sounds & \cite{houben2019foregrounding,houben2020role,houben2022designing} & 3 \\
    ~~Robotic Technology & \cite{machesney2014gerontechnology,lotfi2017active,cruz2018towards,yamazaki2019conversational,cooney2020exercising} & 5 \\
    \hdashline[0.5pt/5pt]
    Others & \cite{bellini2010innovative,morrissey2015creative,lazar2017supporting,perugia2020engage,rik2022access} & 4\\
    \hline
     \textbf{Caregiver Support} & & \textbf{16} \\
     \hdashline
     Assistive Technology & \cite{zachos2013software,maiden2013computing,webster2014technology,homdee2019agitation,alam2019multiple,gomes2020healing} & 6\\
     Robotic Technology & \cite{de2018users,moharana2019robots,carros2020exploring,lv2020teleoperation,yuan2022robot, lee2023reimagining} & 6 \\
     Caregiver Training & \cite{boyd2014investigation,foong2017vita,hiramatsu2020development,shen2023dementia} & 4 \\
     
    \hline
    \textbf{Cognitive Support} & \cite{berenbaum2011augmentative,zaccarelli2013computer,unbehaun2018mobiassist,alves2018enabling,caggianese2018towards,wolf2019care,van2020relivring,maddali2022investigating,spalla2022designing} & \textbf{9} \\
     \hline

    \end{tabular}
\end{table}

\subsubsection{Technologies for Daily Life Monitoring}
% From the selected relevant paper, there are many research about daily life monitoring, including sensors-based technologies~\cite{paletta2018playful}, wearable device~\cite{nesbitt2018reliability}, fall detection~\cite{wang2016location}, safety monitoring~\cite{geddes2010cloud}, etc. These technologies can provide the daily life monitoring for PwD. In generally, PwD's wandering is a challenge for caregivers, and it is not easy to find them. GPS and GPS embedded device or application~\cite{sposaro2010iwander} can offer caregivers an alternative and these tracking system can assist with caregivers and locate the PwD's postion. In certain context, fall detection of PwD are also quite important and sensor-based technology can accurately detect PwD's fall and give them some helps~\cite{~\cite{wang2016location}}. Furthermore, remote monitoring are also play an important role in assist with PwD~\cite{wang2010monitoring}. In addition, sleeping behavior monitoring are also quite important and sensor-based technology can provide some helps with sleeping pattern monitoring~\cite{wang2010monitoring}. 

From the collection of relevant papers, a substantial body of research is dedicated to the domain of "Daily Life Monitoring."
% This category encompasses a wide array of technologies, including sensor-based systems~\cite{paletta2018playful}, wearable devices~\cite{nesbitt2018reliability}, fall detection mechanisms~\cite{wang2016location}, and safety monitoring solutions~\cite{geddes2010cloud}. 
This category encompasses a wide array of technologies, including mobile-based monitoring, sensor-based monitoring, sleeping monitoring, application monitoring, application monitoring and wandering management.
These technologies collectively aim to provide comprehensive monitoring of the daily lives of PwD. Notably, wandering behavior among PwD presents a significant challenge for caregivers, as locating them can be a daunting task. In response to this challenge, the integration of GPS and GPS-embedded devices or applications, as explored in the research, such as "iWander: An Android application for dementia patients" by Sposaro et al.~\cite{sposaro2010iwander}, has emerged as a viable solution. These tracking systems offer an alternative means for caregivers to locate PwD and ensure their safety.
Moreover, the accurate detection of falls among PwD is of paramount importance. Sensor-based technologies, as exemplified in the study "Location-aware fall detection for the elderly" by Wang et al.~\cite{wang2016location}, have proven to be effective in promptly identifying falls and providing timely assistance to individuals in need.
In addition to these applications, remote monitoring technologies, explored in "Continuous remote monitoring of activities of daily living for health independent elderly" by Wang et al.~\cite{wang2010monitoring}, play a pivotal role in supporting PwD by offering caregivers and healthcare professionals the ability to remotely assess and respond to changes in daily activities and routines.
Furthermore, monitoring sleeping behaviors is also a crucial aspect of dementia care. Sensor-based technology, as discussed in the research~\cite{wang2010monitoring}, can provide valuable insights into sleeping patterns, enabling caregivers to better understand and address sleep-related challenges faced by PwD.
These technologies collectively contribute to a comprehensive framework for daily life monitoring, enhancing the quality of care and safety for individuals living with dementia.

\subsubsection{Technologies for Daily Life Support}
% There are also enough innovative technologies about daily life support for dementia patients. These technologies can offer the assistance when they need. For example, the reminder function can be integrated into mobile application and this function can give some reminder in some important things for PwD~\cite{armstrong2010developing}. Smart home can give some assistive support for PwD and this technology can reduce the informal caregiver's work~\cite{hwang2012using}. Furthermore, the other technologies for daily life support, such as remote smart kitchen~\cite{armstrong2010developing},  automatic task assistance~\cite{peters2014automatic}, etc, can provide the assistance for PwD effectively. Although there are not too much research about this direction from the table~\ref{tab:catergories_dementia_care} and Figure~\ref{fig:categories_technology}, technologies in daily life support can extremely help dementia patients solve the daily life stuffs efficiently and enhance their quality life.  

In addition to daily life monitoring, there is a notable body of research focused on "Daily Life Support" technologies tailored to assist individuals with dementia when they need it most. These technologies, such as mobile applications, smart homes, assistive robots, and other assistive technology, offer valuable assistance and support to enhance the daily lives of PwD in various ways.
One noteworthy technology in this category is the integration of reminder functions into mobile applications, as exemplified in the research conducted by Armstrong et al.~\cite{armstrong2010developing}. These reminders play a vital role in helping PwD remember important tasks and events, offering cognitive support and promoting independent living.
Smart home technology also emerges as a promising avenue for daily life support. In some studies, such as a research from Hwang et al.~\cite{hwang2012using}, smart homes are shown to provide assistive support to PwD by automating tasks and creating a safe and comfortable living environment. This technology not only benefits PwD but also alleviates the burden on informal caregivers.
Furthermore, advancements, such as remote smart kitchens~\cite{armstrong2010developing} and automatic task assistance~\cite{peters2014automatic} further contribute to daily life support for PwD. These technologies offer practical assistance in meal preparation and task completion, allowing PwD to maintain a higher level of independence and quality of life.
While the quantity of research in this direction may not be as extensive as some other categories, daily life support technologies play a pivotal role in addressing the daily challenges faced by PwD. Their ability to efficiently assist with daily tasks enhances the overall well-being and autonomy of individuals living with dementia.

% \subsubsection{Technology facilitating social interaction and supporting communication}
\subsubsection{Technologies for Social Interaction and Communication}
%Categories (social interaction): Technologies for Memories, Technologies for Digital Sharing, Technologies for Social Interaction (in Care Settings), Technological Companionships (“pets”, robots, and CUIs)
%Categories (Communicationn): Techonologies facilitating communication between pwd and others, technoloies as interlocutors, technologies for assisting in communication

People with dementia often face communication challenges and experiences of loneliness and social isolation. This poses significant challenges to individuals with dementia, adversely impacting their overall quality of life. Several of the papers in our corpus regarded technologies that could facilitate social interaction and communication for people with dementia, including innovative technologies such as conversational and social robots, e.g., \cite{abdollahi2017pilot}, voice assistants, e.g., \cite{zubatiy2021empowering}, and specialized applications, e.g., \cite{van2020relivring}. Munoz et al. \cite{munoz2021evaluating} explored how a cooperative game could facilitate for better visitations between PwD living in a care home and their visitors. RelivRing \cite{van2020relivring} was developed by researchers to let individuals with dementia relive positive experiences of visits from relatives by listening to audio messages left by their loved ones. "Moments," a system enabling individuals with dementia to engage in artistic expression and online sharing was reported to facilitate a sense of personhood and autonomy \cite{lazar2017supporting}. 

Robotic technology has shown promising results for enhancing social interaction and communication. A robot offering visual and acoustic stimulation in a walking group for people with dementia, has shown promise in enhancing motivation, group coherence, and mood within the group \cite{hebesberger2016lessons}. Robot pet therapy has demonstrated improved moods in elderly individuals \cite{shibata2012therapeutic}, and virtual pet companions on tablets have been employed to reduce loneliness among PwD \cite{machesney2014gerontechnology}. Researchers have specifically explored how robot technology can promote conversations in PwD, for example Yamazaki et al. \cite{yamazaki2019conversational} who demonstrated how a teleoperated android robot could improve behavioral and psychological symptoms of dementia by promoting conversations, and Abdollahi et al. \cite{abdollahi2017pilot} who reported positive results from incorporating a life-like conversational robot into the life of people with moderate dementia and/or depression living in a senior living facility. They found that the robot was a promising tool, but it could not replace human conversations.

% \subsubsection{Technology Enhancing Well-being}
\subsubsection{Technologies for Well-Being Enhancement}
In this section, we focus on technologies designed to enhance the subjective well-being of individuals with dementia by providing meaningful and enriching experiences. These technologies can be divided into two sub groups, where the first focus on technologies used to evoke memories for therapeutic reasons, while the second focus on technologies aimed at creating new and valuable experiences. 
%Categories for well-being/meaningful experiences: Technologies for Therapy, Technologies for Enhancing Everyday Experiences, Technology for Cognitive and Physical Excercise.

Reminiscence therapy (RT), a form of cognitive stimulation therapy centered on evoking memories through discussions and interactions with physical objects, has exhibited promising results in enhancing the well-being of PwD \cite{o2018reminiscence, villasan2021improvement}. The primary goal of RT is to establish a connection between the individual and their past, fostering a renewed sense of personhood \cite{o2018reminiscence}. Conversational and social robots often encompass a variety of tasks, with RT therapy or RT-related activities being one of them. Research in this domain suggests that individuals with dementia generally enjoy interacting with social robots \cite{abdollahi2017pilot}, and that robots specifically designed for RT have led to positive user experiences and fruitful interactions \cite{yuan2023cognitive}. The use of meaningful physical objects as memory triggers is commonly used in RT therapy. Such memory triggers has been explored in the field of HCI, for example through digital photo albums \cite{edmeads2019designing} and photo frames \cite{filoteo2018evaluation}, wall-display \cite{sas2020supporting}, and VR \cite{reisinho2022systematic}. In addition, the integration of sounds to evoke memories and enhance communication has been a subject of exploration in various studies. Everyday sounds have been presented through personalized sound players \cite{houben2022designing}, through the manipulation of familiar objects on a soundboard to trigger corresponding sounds \cite{houben2019foregrounding}, and even through pillow-like sound players designed for individuals with advanced dementia \cite{houben2020role}. These sound-facilitating technologies have shown promise in evoking memories, catering to both early to mid-stage dementia \cite{houben2020role, houben2022designing} and advanced dementia \cite{houben2019foregrounding}. Building on concepts from tangible objects, interactive furniture, such as the Resonant Interface Rocking Chair by Bennett et al. \cite{bennett2016rekindling}, aims to create environments that subtly engage individuals with dementia in activities that stimulate memories and storytelling, often through the activation of familiar sounds. Initial results are promising, with participants enjoying their interaction with the furniture and recalling memories. Other interactive media experiences, such as AmbientEcho \cite{thoolen2020ambientecho}, integrate various technologies like virtual windows, photo frames, and ambient lights and music, designed for care home settings. Personalized and generic content is displayed through these mediums, leading to meaningful engagements among individuals themselves, between residents, and with relatives and care practitioners.

Another means of enhancing the well-being of individuals with dementia involves using technologies, such as VR and social robots, to promote physical and cognitive exercise. Exergames can serve as medium for both cognitive and physical exercise and employ technologies like screen-based video games \cite{lopez2018acceptance, unbehaun2020social, unbehaun2018exploring} and VR experiences \cite{eisapour2018participatory, karaosmanoglu2021lessons}. Designed with specific goals in mind, these games aim to promote physical activity \cite{eisapour2018participatory, unbehaun2020social} or encourage cognitive stimulation \cite{lopez2018acceptance, karaosmanoglu2021lessons}, ultimately contributing to the overall well-being of individuals with dementia. In addition to games, social robots have also be employed to engage individuals with dementia in exercise therapy, helping them maintain their motivation \cite{cruz2018towards}. Furthermore, autonomous social robots have demonstrated their capability to facilitate cognitive stimulation therapy, resulting in improved quality of life for patients \cite{cruz2020social}.

%New Experiences in Everyday Life
While some technologies have clear and specific therapeutic purposes, others have more open and ambiguous goals, centered around creating meaningful experiences for individuals with dementia. These technologies aim to enrich the lives of people with dementia through novel experiences. Innovative approaches for tangible objects, such as those explored by Houben et al. \cite{houben2022enriching}, seek to "promote social participation and pleasurable experiences in everyday care situations" through tangible objects. Many individuals with dementia reside in care facilities, often constrained by physical limitations and the availability of care resources. These constraints limit the range of experiences they can access. VR presents an innovative avenue for introducing individuals with dementia to new experiences, potentially improving their overall well-being. For instance, in the study conducted by Tabbaa et al. \cite{tabbaa2019bring}, researchers developed a VR system \cite{matsangidou2022free} and tested it with individuals with moderate to severe dementia in a locked psychiatric hospital. Participants reported numerous benefits, including the sense of having a personal space and the opportunity to explore environments beyond their usual surroundings. These experiences contributed to improved mood and overall well-being \cite{tabbaa2019bring}. Similarly, in another approach, researchers have transformed physical spaces within care settings to create multisensory experiences for residents at advanced stages of dementia \cite{gomes2020healing}. This work involved the integration of both physical and digital components to craft sensory-rich experiences. The primary goal of these experiences was to "transport" users to settings capable of evoking feelings of mindfulness and serenity often associated with natural environments. Initial results have been highly positive, benefiting both residents and caregivers.

\subsubsection{Technologies for Caregiver Support}
% The technologies for caregiver support mainly focuses on developing solutions and tools to provide assistance, education, and emotional support to the informal caregivers responsible for the well-being of individuals with dementia.
% These technologies can alleviate some of the burdens while enhancing the quality of care. In our literature review, there are not too much paper about this category from the Table~\ref{tab:catergories_dementia_care}, although assisting caregivers can also improve dementia care. Caregivers' training can offer informal caregivers access to educational materials, training modules, and resources to better understand dementia, its progression, and effective caregiving techniques. Online course~\cite{boyd2014investigation} and AR-based dementia education~\cite{shen2023dementia} can effectively educate caregivers and improve their care skills, and then enhance the dementia care for PwD. In addition, caregivers can utilize technology to remotely monitor the well-being of individuals with dementia~\cite{homdee2019agitation,alam2019multiple}. The collected data from remote sensors or wearable devices can effectively help caregiver know the status of PwD and also offer the appropriate healthcare. 
% Moreover, emotion support and well-being for caregivers are also quite important although there is no paper in our literature review about this research. Technology can provide emotional support to caregivers through certain mindfulness and stress reduction apps, and then caregivers can provide a good dementia care for PwD, especially for formal caregivers.
The category of caregiver support technologies is primarily focused on the development of solutions and tools aimed at providing essential aid, education, and emotional sustenance to informal caregivers responsible for the well-being of individuals with dementia. 
This category includes assistive technology, robotic technology and caregiver training.
Although this category was relatively underrepresented in our literature review, its potential to alleviate caregiver burdens and enhance the quality of dementia care cannot be understated.
There are not many papers in our evaluation regarding this category from the Table~\ref{tab:catergories_dementia_care}, however aiding carers can also enhance dementia care.
The technologies in education for caregivers can empower caregivers by granting them access to educational materials, training modules, and resources designed to deepen their understanding of dementia, its progression, and effective caregiving techniques. Examples include online courses~\cite{boyd2014investigation} and innovative approaches like AR-based dementia education~\cite{shen2023dementia}. Such tools effectively educate caregivers and improve their caregiving skills, thereby enhancing the quality of dementia care provided to individuals with dementia.
Additionally, caregivers can leverage technology for remote monitoring of individuals with dementia, as demonstrated in some studies\cite{homdee2019agitation, alam2019multiple}. These technologies employ remote sensors and wearable devices to provide caregivers with real-time insights into the well-being and vital signs of those under their care. This capability enables timely healthcare interventions and ensures the safety and health of individuals with dementia.
While our literature review may not have yielded papers specifically focused on emotional support for caregivers, it is vital to recognize the significance of this aspect. Technology has the potential to provide emotional sustenance to caregivers through mindfulness and stress reduction apps, fostering their mental and emotional well-being. This emotional support enables caregivers, especially formal caregivers, to provide higher-quality dementia care.
To summarize, the development of technologies for caregiver support represents a crucial aspect of enhancing dementia care. These solutions offer caregivers the knowledge, tools, and emotional resilience needed to provide the best possible care for individuals with dementia. Although the volume of research in this category may be limited, its impact on improving the overall caregiving experience and the well-being of both caregivers and those they care for is substantial.

\subsubsection{Technologies for Cognitive Support}
The primary objective of cognitive support technology is to preserve cognitive function, slow down the progression of cognitive decline, and enhance the overall quality of life for individuals with dementia. These technologies empower people with dementia to lead more independent and fulfilling lives, delivering benefits not only to formal caregivers but also to family members involved in their care.
While the number of papers in the "Cognitive Support" category may be relatively small from the Table~\ref{tab:catergories_dementia_care}, the importance of technologies designed for cognitive support cannot be overstated.
Among these technologies, VR/AR-based solutions have gained prominence in recent years, as evidenced by research~\cite{caggianese2018towards,wolf2019care,spalla2022designing}. These VR/AR environments offer a safe and controlled space for individuals with dementia to engage in cognitive rehabilitation exercises, thereby promoting their independence within a home setting.
Additionally, mobile applications and software play a significant role in cognitive support. For instance, SAVION is a software program designed to provide cognitive stimulation and training for individuals with dementia~\cite{berenbaum2011augmentative}. RelivRing, another application, allows individuals with dementia to relive cherished memories by listening to audio messages left by their loved ones~\cite{berenbaum2011augmentative}.
Further avenues in cognitive support encompass diverse approaches to cognitive therapy. Information and communication technology (ICT)-based training systems~\cite{unbehaun2018mobiassist}, and robot-based adaptive cognitive stimulation systems~\cite{ros2020exploration}, offer innovative and tailored methods for supporting cognitive health. Overall, cognitive support technologies hold great promise in improving the lives of individuals with dementia. They provide essential tools and interventions that promote cognitive well-being, enhance independence, and enrich the overall dementia care experience for both formal caregivers and family members.
\section{Discussion}
% This paper present ongoing research on existing technologies in dementia care. Comparing to the conventional dementia care, current technologies can provide many support for dementia patients and their caregivers, such as daily life monitoring based sensor-based technologies or wearable devices. Moreover, these technologies can also effectively help caregivers with their dementia care and improve dementia patients and their caregivers' well-being. It means these technologies can build a bridge between traditional dementia care and PwD and their caregivers. 
% Through analyzing the final selected paper, our results confirmed technologies in dementia care has become a hot research direction, especially recent years. Based on the research from the selected paper, we divided these technologies in dementia care into six categories, and \textit{well-being enhancement}, \textit{social Interaction and Communication} and \textit{Daily Life Support} became currently main research area from our findings. This finds guide us to the future research in dementia care. In this section, we will give some discussion and analysis based our results. 
This paper presents a comprehensive overview of ongoing research in the field of dementia care technology, highlighting its potential to significantly enhance the lives of dementia patients and their caregivers compared to conventional care approaches. The advancements in technology have introduced a range of supportive solutions, including daily life monitoring through sensor-based technologies and wearable devices. These technologies not only benefit dementia patients by improving their quality of life but also offer substantial assistance to caregivers, ultimately bridging the gap between traditional care and the unique needs of both patients and caregivers. The analysis of selected papers firmly underscores the increasing prominence of technology in dementia care research, particularly in recent years. From our findings, we have categorized these technologies into six distinct categories, with \textit{Well-being Enhancement}, \textit{Social Interaction and Communication} and \textit{Daily Life Monitoring} emerging as the primary research areas. These findings offer valuable guidance for future research in dementia care.
In this section, we engaged in a thoughtful discussion and analysis based on our findings.

%Limited results on the different technologies, due to them being innovate/new. Things may change, other findings may appear. 

%The focus in most of our data
\subsection{Consideration of Dementia Progression}
While many technologies aim to enhance the lives of people with dementia, such as intelligent assistive technologies~\cite{dada2021intelligent}, mobile-based solutions~\cite{koo2019examining,ye2023researched}, ambient-assisted living technologies~\cite{gettel2021dementia}, and smart health technologies~\cite{guisado2019factors},etc., they often predominantly focus on individuals with later onset dementia, typically diagnosed after the age of 65. This focus leaves a notable gap in research and care for those with early onset dementia, also known as young onset dementia. Early onset dementia affects individuals who are diagnosed at a younger age, and addressing their specific needs and challenges is of paramount importance \cite{yuan2023cognitive}. 
% Unlike older adults, dementia in young people is often misdiagnosed or overlooked due to its rarity and Early diagnosis is crucial for accessing appropriate support services and interventions~\cite{sullivan2022peer}.
Due to its rarity, dementia in young individuals is frequently misdiagnosed or ignored, in contrast to older adults~\cite{sullivan2022peer}. Early diagnosis is essential for obtaining appropriate support services and therapies.
Younger individuals may have different care needs compared to older adults with dementia~\cite{hancock2006needs}. They may require services tailored to their age group, such as vocational rehabilitation, educational support, and assistance with maintaining social connections. 
% Further, young people with dementia often want to maintain their independence and sense of identity for as long as possible~\cite{greenwood2016experiences}. Support services should focus on enabling them to remain active and engaged in activities they enjoy, while also providing assistance with tasks they find challenging.
Furthermore, young individuals with dementia frequently want to preserve their independence and sense of self for as long as feasible~\cite{greenwood2016experiences}. Support services should focus on keeping people active and involved in things they love, as well as assisting them with difficult chores.
However, some researchers have underscored the importance of addressing the unique requirements of individuals with early onset dementia and advocated for the implementation of dementia care technologies tailored to this demographic \cite{yuan2023cognitive}. The distinct experiences and challenges faced by individuals diagnosed at a younger age necessitate innovative approaches to technology-driven care and support.

Moreover, it is essential to recognize that dementia manifests across various stages, each of which presents different cognitive, physical, and emotional challenges. The progression of dementia can significantly influences how individuals interact with technology, and this aspect deserves further exploration. While technology has the potential to improve the quality of life for PwD at different stages of dementia, it should be designed with an understanding of the unique needs and capabilities associated with each stage.
Incorporating technology into dementia care for early onset cases and tailoring solutions to different stages of dementia will not only address existing gaps in research but also contribute to more effective and compassionate care for individuals across the entire dementia spectrum.

%Research on conversational technologies for PwD without or with limited language

\subsection{Emerging Technologies and Future Trends}

Research in the field of HCI concerning technology for PwD has primarily emphasized enhancing well-being in recent years, as shown in Figure~\ref{tab:catergories_dementia_care}. This underscores the growing importance of technologies aimed at improving the quality of life for both dementia patients and their caregivers.
In addition to conventional dementia care approaches, innovative HCI technologies like virtual reality and mindfulness applications are emerging as valuable tools to enhance the well-being of individuals with dementia. These technologies offer promising avenues for improving the overall quality of life for this demographic.
The recent advancements in sensors and wearable technologies have brought about significant developments in both the \textit{Daily Life Monitoring} and \textit{Daily Life Support} categories. Smart homes and assistive devices \cite{lazarou2016novel}, among other solutions, are simplifying daily routines for dementia patients while simultaneously alleviating the burden on caregivers. These practical aids hold the potential to greatly enhance the lives of both patients and their caregivers.
Moreover, the continuous progress in large language models and natural language-based technologies suggests a potential increase in research focused on social interactions and communication. These innovations may not only bolster social robots and communication apps but also extend their application to other categories detailed in Figure~\ref{tab:catergories_dementia_care}. These technologies can offer companionship and emotional support to individuals with dementia, ultimately reducing the physical and emotional strains on caregivers. 
% For example, personalised voice assistant for dementia patients and caregivers \cite{li2020personalized} and social assistant robot for people with dementia can give the helps for dementia patients. 
For instance, a personalized voice assistant tailored for dementia patients and caregivers ~\cite{li2020personalized} and social assistant robots designed for people with dementia~\cite{striegl2021designing} offer valuable assistance in managing the dementia and enhancing overall well-being.

Currently, the ongoing research in dementia care technology holds enormous promise for improving how we support individuals living with dementia. 
For instance, the innovative Rewind system leverages self-tracked location data to aid in recalling everyday memories~\cite{tan2018rewind}. While currently not implemented in dementia care, the potential benefits for future applications in this field are promising.
Additionally, NeuralGait represents another noteworthy smartphone-based system designed to passively capture gait data for assessing brain health~\cite{li2023neuralgait}. Its potential application in future dementia care holds significant promise.
By addressing their unique needs and those of their caregivers, these technologies have the potential to revolutionize dementia care and improve the well-being of those affected by this condition. As we continue to advance in this field, it is crucial to maintain a strong focus on ethical considerations, longitudinal studies, and real-world implementation to ensure that these technologies deliver on their promise and truly enhance the lives of dementia patients and their caregivers.

\subsection{Limitation}
% In this literature review work, we only focus on the database of ACM and IEEE from 2010 to 2023, and this may lead to some bias in the final result. Although our findings show that current ongoing research regarding technologies on dementia care are mainly focus on \textit{well-being enhancement}, \textit{social Interaction and Communication}, \textit{Daily Life Monitoring} and \textit{Daily Life Support} are not important research in the future direction. 
% In the other words, the technologies on these two categories may play a quite important role in the future innovative technologies in dementia care. while cognitive support in dementia care may have received less attention in recent research, it plays a crucial role in improving the lives of PwD.
% Innovative technologies like VR/AR, mobile applications, and cognitive training systems have the potential to enhance cognitive abilities, promote independence, and contribute to the overall well-being of individuals with dementia. Future research in this area holds promise for further advancements in dementia care. 
% In addition, caregiver assistance in dementia care is an essential area of research and development. By providing caregivers with the necessary tools, education, and emotional support, technology can empower them to deliver higher-quality care and improve the overall well-being of both PwD and caregivers. Future research in this field should continue to explore innovative solutions to better support caregivers in their important role.
In this literature review, we acknowledge the limitations of our focus on the ACM and IEEE databases from 2010 to July 2023. This approach may introduce some bias in our findings, as it emphasizes certain research directions over others. While our results highlight the prominence of research on technologies aimed at \textit{Well-being Enhancement}, \textit{social Interaction and Communication},and \textit{Daily Life Monitoring}, we must recognize the importance of the other two categories in future dementia care.
Indeed, technologies in \textit{Cognitive Support} and \textit{Daily Life Support} are critical components of dementia care. With advancements in sensors and wearable devices, these technologies hold great potential to simplify daily routines for individuals with dementia and alleviate the burden on caregivers. Therefore, they remain valuable areas for future research and innovation.
Furthermore, \textit{Cognitive Support} in dementia care, despite receiving comparatively less attention in recent research, plays a pivotal role in enhancing the lives of individuals with dementia. Innovative technologies such as VR/AR applications, mobile cognitive training tools, and adaptive cognitive stimulation systems offer promising avenues for improving cognitive function, promoting independence, and enhancing overall well-being among people with dementia. Future research in this domain holds significant promise for advancing dementia care.
Lastly, \textit{caregiver support} in dementia care stands as an essential field of research and development. Equipping caregivers with the necessary resources, training, and emotional support through technology can empower them to deliver higher-quality care. This not only benefits individuals with dementia but also improves the overall well-being of caregivers themselves. Future research should continue to explore innovative solutions to better support caregivers in their indispensable role.

\section{Conclusion}
In this work, we have delved into the landscape of dementia care technologies by analyzed literature published in ACM and IEEE from 2010 to 2023. Our literature review approach, guided by the PICO process and PRISMA guidelines, led us to a final selection of 133 relevant papers. These papers shed light on the current state of technologies designed to enhance dementia care.
Our exploration of these selected papers allowed us to categorize existing dementia care technologies into six distinct categories: \textit{Daily Life Monitoring }, \textit{Daily Life Support}, \textit{Social Interaction and Communication}, \textit{Well-being Enhancement}, \textit{Cognitive Support}, \textit{Caregiver Support}. Each category represents a vital aspect of dementia care, addressing the diverse needs of both patients and caregivers.
Furthermore, we have provided a glimpse into the research agenda for each of these categories, showcasing the ongoing advancements and innovations in the field. Wearable devices, mobile applications, robotic technologies, and more have been at the forefront of research, offering promising solutions to improve the lives of those affected by dementia.
Looking ahead, we have discussed potential future technologies that could revolutionize dementia care. These technologies, ranging from wearable devices to Human-Computer Interaction (HCI)-driven solutions, hold the promise of enhancing the quality of life for dementia patients and alleviating the burdens faced by caregivers.
In conclusion, our work serves as a valuable guide for future research in dementia care. By researching the wealth of knowledge within this critical domain, we hope to inspire and inform researchers, caregivers, and healthcare professionals alike, ultimately advancing the care and support available to individuals living with dementia.

% \begin{acks}
% To be decide
% \end{acks}

\bibliographystyle{ACM-Reference-Format}
\bibliography{reference}

\appendix





\end{document}